\documentclass[12pt]{article}
\usepackage{authblk}
\usepackage[bookmarksnumbered, colorlinks, plainpages]{hyperref}
\usepackage{amsmath, amsthm, amscd, amsfonts, amssymb, graphicx, color, booktabs}
\numberwithin{equation}{section}
\definecolor{email}{rgb}{0.00,0.00,0.84}
\begin{document}
\setcounter{page}{1}

\title{\large \bf 12th Workshop on the CKM Unitarity Triangle\\ Santiago de Compostela, 18-22 September 2023 \\ \vspace{0.3cm}
\LARGE Radiative $B$ decays at Belle and Belle II}

\author[1]{Rahul Tiwary on behalf of the Belle and Belle II Collaborations}
\affil[1]{Tata Institute of Fundamental Research, Mumbai 400005, India}
\maketitle

\begin{abstract}
Rare decays of $B$ mesons to radiative final states serve as an ideal ground to search for New Physics effects from short range contributions. Observed by CLEO in the early 1990s as one of the first transitions of the electroweak penguin family, these decays continue to play a key role in testing predictions of the Standard Model. We present the latest results from Belle and Belle II for the radiative penguin transitions.
\end{abstract} \maketitle

\section{Introduction}

\noindent 

In the Standard Model (SM), transitions involving Flavour Changing Neutral Currents (FCNC) are forbidden at the tree level~\cite{GIM}, and occur through electroweak loop diagrams. Beyond-the-SM (BSM) physics can either contribute to the loop diagrams or directly appear at tree level, which would alter various physics observables from their SM predictions. 

A typical FCNC transition is the radiative decay of a $B$ meson to the inclusive final state  $X_{s}\gamma$, where $X_{s}$ includes all final states with net strangeness~\cite{R1}. This decay offers excellent sensitivity to BSM effects~\cite{R2}. The photon-energy spectra can also provide insights into SM parameters, such as the $b$ quark mass and its motion within the $B$ meson~\cite{R3, R4}. BSM searches using radiative decays of $B$ mesons to exclusive final states, like $B \to K^{*}\gamma$ and $B \to \rho\gamma$, are also promising. The $b \to s/d\gamma$ operator is the dominant contributor to the decay, making it easier to tell the difference between SM and BSM physics contributions~\cite{Belle_II_Physics}. 

The remainder of this document is arranged as follows: Section~\ref{Detector} provides a brief description of Belle and Belle II detectors; Section~\ref{BtoXsgamma} summarizes the measurement of photon energy spectrum obtained for the inclusive $B\to X_{s}\gamma$ channel using Belle II data; and Section~\ref{Btorhogamma} provides a summary of the measurement of observables of exclusive final state $B\to \rho \gamma$ using the combined Belle and Belle II dataset. 

\section{The Belle and Belle II detectors}
\label{Detector}
\noindent
The Belle detector~\cite{Belle_TDR, Belle_Physics} was a large-solid-angle spectrometer that operated at the KEKB asymmetric-energy $e^{+}e^{-}$ collider~\cite{R8, R9}. The energies of the electron and positron beams were 8.0 GeV and 3.5 GeV, respectively. The detector consisted of a silicon-strip vertex detector, a central drift chamber, an array of aerogel Cherenkov counters and
time-of-flight scintillation counters for identification of
charged particles, and a CsI(Tl)-based electromagnetic calorimeter (ECL), all of which were surrounded by a superconducting solenoid coil providing a magnetic field of 1.5\,T. An iron flux return yoke located outside the coil instrumented with resistive-plate chambers to facilitate the detection of $K^{0}_{\rm L}$ mesons and to identify muons. 

Belle II~\cite{Belle_II_TDR} is an upgraded version of Belle and located at the SuperKEKB~\cite{SuperKEKB} $e^{+}e^{-}$ collider. The energies of electron and positron beams are 7.0\,GeV and 4.0\,GeV, respectively. The Belle II detector includes two layers of silicon pixel sensors, four layers of double-sided silicon-strip vertex detectors~\cite{R11} and an upgraded 56-layer central drift chamber. Two types of Cherenkov-light detector systems surround the drift chamber: an azimuthal array of time-of-propagation detectors for the barrel region and an aerogel ring-imaging Cherenkov detector for the forward endcap region. Belle II reuses the ECL of Belle along with its solenoid and the iron flux return yoke; the latter is equipped with both resistive-plate chamber and plastic scintillator modules to detect $K^{0}_{\rm L}$ mesons and muons. The $z$ axis of the laboratory frame is defined as the solenoid axis, where the positive direction is along the electron beam. This convention applies both to Belle and Belle II.

\section{\texorpdfstring{Inclusive measurement of $B \to X_{s}\gamma$ at Belle II}{Inclusive measurement of B -> Xs gamma at Belle II}}
\label{BtoXsgamma}
\noindent 
This section summarizes a measurement of inclusive $B \to X_{s}\gamma$ decays using hadronic tagging for the partner $B$ meson ($B$ tag candidate) reconstruction, performed at Belle II~\cite{BtoXsgamma_BelleII}. The measurement is based on $189\,\rm {fb}^{-1}$ of data collected by the Belle II experiment. This approach is complementary to the untagged or lepton-tagged~\cite{R12} and sum-of-exclusive~\cite{R13} methods. The hadronic tag ensures higher signal purity, whereas the kinematic constraints from the $B$ tag candidate allows us to measure observables pertaining to the signal-side $B$ meson. 

To reconstruct a $B \to X_{s}\gamma$ candidate, we pair the highest energy photon in the event, satisfying $E^{B}_{\gamma} > 1.4$\,GeV, with the $B$ tag candidate reconstructed using the full-event-interpretation (FEI) algorithm~\cite{R14}. Here, the photon energy in the rest frame of the signal-side $B$ meson is denoted as $E^{B}_{\gamma}$. The FEI algorithm reconstructs hadronic $B$ decays from numerous sub-decay chains. We use a stochastic gradient-boosted decision tree (BDT)~\cite{BDT}, trained on Zernike moments~\cite{Zernike} to separate high-energy photons from $K^{0}_{\rm L}$ clusters.

The sources of background include high-energy or hard photons coming from $\pi^{0}/\eta \to \gamma\gamma$ decays, where a photon emitted along the boost direction of $\pi^{0}/\eta$ can be misidentified as the signal-side photon. Another contribution comes from $e^+e^- \rightarrow q\overline{q}$ decays, where $q$ represents $u, d, s,$ and $c$ quarks. The latter is also referred to as continuum background. To mitigate the background from $\pi^0$ and $\eta$ photons, the signal-side hard photon is paired with all low-energy photons in the event. Events having pairs consistent with $\pi^0/\eta \to \gamma\gamma$ decays are then vetoed using a dedicated BDT trained on kinematic variables such as the diphoton invariant mass, helicity, and properties of the low-energy photon, including its energy, polar angle, and smallest cluster-to-track distance. Another dedicated BDT trained to suppress continuum background is employed. The BDT utilizes input features that are known to have good separation power between $B$ decays and continuum background; these features include the modified Fox-Wolfram moments \cite{KSFW}, CLEO cones \cite{CLEO}, thrust, etc. 

A fit to the $M_{\rm bc} (\equiv \sqrt{(E^{\ast}_{\rm beam})^{2}-(\vec{p}^{\,\ast}_{B})^{2}})$ distribution of $B$ tag candidates in bins of $E^{B}_{\gamma}$ for the region $E^{B}_{\gamma}>1.8$\,GeV is performed to extract the signal yield. Here, $E^{\ast}_{\rm beam}$ and $E^{\ast}_{B}$ are the beam energy and $B$-meson energy in the center-of-mass frame. The selections and fit procedures are validated in a control region $1.4 < E^{B}_{\gamma} < 1.8$\,GeV. Since this inclusive analysis does not differentiate between $b \to d\gamma$ and $b \to s\gamma$ processes, we subtract the smaller $b \to d\gamma$ contribution based on simulation. The measured $B \rightarrow X_{s}\gamma$ spectrum undergoes correction (unfolding) for smearing effects. The unfolding procedure uses bin-by-bin multiplicative factors, obtained as the ratios between the expected number of events in the generated spectrum and those in the reconstructed spectrum within a given $E^{B}_{\gamma}$ interval. The integrated branching fractions for various $E^{B}_{\gamma}$ thresholds are listed in Table~\ref{tab:BtoXsgamma_BF}, the results are consistent with world average values~\cite{PDG}.

\begin{table}[htb!]
\caption{The integrated partial branching fractions for three $E^{B}_{\gamma}$  thresholds.}\label{tab:BtoXsgamma_BF}
\renewcommand\arraystretch{1.5}
\noindent\[
\begin{tabular}{cc}
    $E^{B}_{\gamma}$thres.\,(GeV) & $\mathcal{B}(B\to X_{s}\gamma)\times(10^{-4})$ \\
    \hline
    1.8 & 3.54 $\pm$ 0.78 (stat.) $\pm$ 0.83 (syst.) \\
     2.0 & 3.06 $\pm$ 0.56 (stat.) $\pm$ 0.47 (syst.)\\
    2.1 & 2.49 $\pm$ 0.46 (stat.) $\pm$ 0.35 (syst.)\\
    \hline 
\end{tabular}
\]
\end{table}

\section{\texorpdfstring{Exclusive measurement of $B\to \rho \gamma$ at Belle and Belle II}{Exclusive measurement of B-> rho gamma at Belle and Belle II}}
\label{Btorhogamma}
\noindent
This section summarizes the most precise measurement of observables for exclusive $B\to \rho \gamma$ decays, based on a combined data sample of the Belle ($711\,\rm {fb}^{-1}$) and Belle II ($362\,\rm {fb}^{-1}$) experiments. The exclusive radiative decay of $B$ mesons to the $\rho\gamma$ final state allows for an independent search of BSM, complementary to the $b\to s\gamma$ modes. The $B \to \rho\gamma$ decays, which is a $b \to d$ quark-level transition, has a branching fraction an order of magnitude smaller than the radiative $B$ decays involving a $b \to s$ transition. Owing to a significant difference in the branching fractions of $b \to d \gamma$ and $b \to s \gamma$ transitions, one needs good particle identification detectors to cut down the charged kaon contamination from $B \to X_{s}\gamma$ decays. The branching fractions give weak constraints on BSM parameters, since their SM predictions suffer from large uncertainties (around 20\%) due to the form factors~\cite{BF}. One can instead study observables such as $C\!P$ ($\mathcal{A}_{C\!P}$) and isospin asymmetry ($\mathcal{A}_{I}$), which are theoretically cleaner due to the cancellation of such effects. Precision measurement of the $\mathcal{A}_{I}$ in $B \to \rho\gamma$ is particularly interesting since the current world average~\cite{PDG} is in slight tension with the SM~\cite{AI}. 
 
The reconstruction of $B\to\rho \gamma$ decay follows a hierarchical approach, starting with the final-state particles. Hard photon candidates exhibiting a shower shape consistent with that of an isolated photon are selected within the energy range of 1.8 to 2.8 GeV. Tracks produced near the $e^{+}e^{-}$ interaction point are selected based on the requirements ${|d_{r}| < 0.5~\rm cm }$ and ${|d_{z}| < 2.0~\rm cm }$, where $d_{r}$ ($d_{z}$) denotes the track's transverse (longitudinal) impact parameters. A likelihood-based particle selector combining information from various detectors of Belle~\cite{PID} or Belle II is used to identify charged tracks.

The $\pi^{0}$ candidates are reconstructed in the diphoton invariant-mass range of $119 < M_{\gamma\gamma} < 151\,{\rm MeV}\!/c^2$. The photons are further required to satisfy various energy thresholds depending on the detector (Belle or Belle II) and the region of the ECL where the photon is detected. Subsequently, we reconstruct the $\rho$ mesons via $\rho^{0} \rightarrow \pi^{+}\pi^{-}$ and $\rho^{+} \rightarrow \pi^{+}\pi^{0}$ modes with the selection 0.64 (0.65) $< M_{\pi\pi} <$ 0.89 (0.90)$\,{\rm GeV}\!/c^2$ for Belle (Belle II). We reconstruct the $B$ meson by combining a high-energy photon with the pion pair. Further selection criteria are applied to the variables $M_{\rm bc}> 5.2\,\rm GeV\!/c^{2}$ and $|\Delta E (\equiv E^{\ast}_{B}-E^{\ast}_{\rm beam} )|<0.3\,{\rm GeV}$. For the neutral mode, the momentum of the $B$ meson in the center-of-mass frame is calculated as
$\vec{p}^{\,\ast}_{B^{0}} = \vec{p}^{\,\ast}_{\rho^{0}} + \frac{\vec{p}^{\,\ast}_{\gamma}}{|\vec{p}^{\,\ast}_{\gamma}|} \times (E^{\ast}_{\rm beam}-E^{*}_{\rho^{0}})$, to improve the resolution of $M_{\rm bc}$.

The sources of background, akin to the $B\to X_{s}\gamma$ channel, are hard photons from $\pi^{0}\!/\eta$ decays, and combinatorial background from $e^+e^- \rightarrow q\overline{q}$ events. The background suppression strategy is to use a dedicated BDT classifier to suppress each kind of background, the same as $B\to X_{s}\gamma$. $B^+ \to D^{0}[K^-\pi^+]\pi^+$, $B^{0} \to D^{-}[K^+\pi^-\pi^-]\pi^+$, $B^{0} \to K^{*0}[K^+\pi^-]\gamma$, and $B^{+} \to K^{*+}[K^+\pi^{0}]\gamma$ control channels are studied to assess quality of the simulation and assign systematics for the BDT classifiers. These control channels have a similar final state as the $B\to \rho \gamma$ modes but with significantly higher statistics, and relatively low background contamination. The invariant mass regions for $D^{0}$ $(D^{-})$ and $K^{*}$ particles are 1.85 (1.86) $< M_{K^-\pi^+} (M_{K^+\pi^-\pi^-}) <1.88\,{\rm GeV}\!/c^2$, and $0.817 < M_{K\pi} < 0.967\,{\rm GeV}\!/c^2$, respectively.

The physics observables of $B\to \rho\gamma$ decay are obtained from an extended maximum-likelihood fit to unbinned $M_{\rm bc}$, $\Delta E$, and $M_{K\pi}$ distributions, performed simultaneously for six independent datasets: $B^+$, $B^-$, and $B^0$ in Belle and Belle II. Here, $M_{K\pi}$ is the invariant mass calculated assuming a $\pi^{+}$ to be a $K^{+}$. Using $M_{K\pi}$ instead of $M_{\pi\pi}$ aids in better separation of the $B \rightarrow K^{*}\gamma$ background. The measured observables for $B\to \rho\gamma$ decays from combined Belle and Belle II datasets are as follows:
\begin{equation}
\mathcal{B}(B^+ \to \rho^+\gamma) = (13.1^{+2.0+1.3}_{-1.9-1.2}) \times 10^{-7},
\end{equation}
\begin{equation}
\mathcal{B}(B^0 \to \rho^0\gamma) = (7.5 \pm1.3^{+1.0}_{-0.8}) \times 10^{-7},
\end{equation}
\begin{equation}
\mathcal{A}_{C\!P}(B \to \rho\gamma) = -8.2 \pm 15.2^{+1.6}_{-1.2} \%,
\end{equation}
\begin{equation}
\mathcal{A}_{I}(B \to \rho\gamma) = 10.9^{+11.2+6.8+3.8}_{-11.7-6.2-3.9} \%.
\end{equation}

The third uncertainty appearing for $\mathcal{A}_{I}$ measurement is due to the ratio of branching fraction of $\Upsilon(4\text{S})$ to charged and neutral $B$ meson pairs. These are the most precise measurements of $B\to \rho \gamma$ observables to date and supersede the previous measurements performed by Belle~\cite{Belle_Btorhogamma}.

\section{Acknowledgment}
The author expresses gratitude to the Belle II Collaboration for the opportunity to present the talk, and commends organizers of the 12th Workshop on the CKM Unitarity Triangle for hosting a successful conference.

\bibliographystyle{amsplain}

\end{document}